\documentclass[final,authoryear,5p]{elsarticle}
\usepackage{graphics}
\usepackage{amssymb}
\usepackage{amsmath}
\usepackage{bm}
\usepackage[colorlinks = true, linkcolor = blue, urlcolor  = blue, citecolor = blue, anchorcolor = blue]{hyperref}
\usepackage{hyperref}

\usepackage{graphicx}
\usepackage{subcaption}
\usepackage{listings}
\usepackage{color}
\usepackage{enumitem}
\usepackage{bibentry}
\nobibliography*

\definecolor{dkgreen}{rgb}{0,0.6,0}
\definecolor{gray}{rgb}{0.5,0.5,0.5}
\definecolor{mauve}{rgb}{0.58,0,0.82}

\journal{Advances in Space Research}

\begin{document}

\begin{frontmatter}

\title{MHD Code Using Multi Graphical Processing Units: SMAUG+}

\author[label1,label3,label4]{N. Gyenge\corref{cor}}
\cortext[cor]{Corresponding author}
\ead{n.g.gyenge@sheffield.ac.uk}
\author[label2]{M. K. Griffiths}
\author[label1,label4]{R. Erd\'elyi}

\address[label1]{Solar Physics and Space Plasmas Research Centre (SP2RC), School of Mathematics and Statistics,\\University of Sheffield, Hounsfield Rd, Sheffield S3 7RH, UK}

\address[label3]{Debrecen Heliophysical Observatory (DHO), Konkoly Observatory, Research Centre for Astronomy and Earth Sciences\\Hungarian Academy of Sciences, Debrecen, P.O.Box 30, H-4010, Hungary}

\address[label2]{Corporate Information and Computing Services, The University of Sheffield, 10-12 Brunswick Street, Sheffield S10 2FN, UK.}

\address[label4]{Department of Astronomy, E\"otv\"os L\'or\'and University, Pf. 32, Budapest, H-1518 Hungary}

\begin{abstract}

This paper introduces the Sheffield Magnetohydrodynamics Algorithm Using GPUs (SMAUG+), an advanced numerical code for solving magnetohydrodynamic (MHD) problems, using multi-GPU systems. Multi-GPU systems facilitate the development of accelerated codes and enable us to investigate larger model sizes and/or more detailed computational domain resolutions. This is a significant advancement over the parent single-GPU MHD code, SMAUG (\bibentry{griffiths2015abs}). Here, we demonstrate the validity of the SMAUG+ code, describe the parallelisation techniques and investigate performance benchmarks. The initial configuration of the Orszag-Tang vortex simulations are distributed among 4, 16, 64 and 100 GPUs. Furthermore, different simulation box resolutions are applied: $1000 \times 1000$, $2044 \times 2044$, $4000 \times 4000$ and $8000 \times 8000$.  We also tested the code with the Brio-Wu shock tube simulations with model size of 800 employing up to 10 GPUs. Based on the test results, we observed speed ups and slow downs, depending on the granularity and the communication overhead of certain parallel tasks. The main aim of the code development is to provide massively parallel  code without the memory limitation of a single GPU. By using our code, the applied model size could be significantly increased. We demonstrate that we are able to successfully compute numerically valid and large 2D MHD problems.

\end{abstract}


\end{frontmatter}

\section{Introduction}

\begin{figure*}[ht]
	\centering
	\includegraphics[width=184mm]{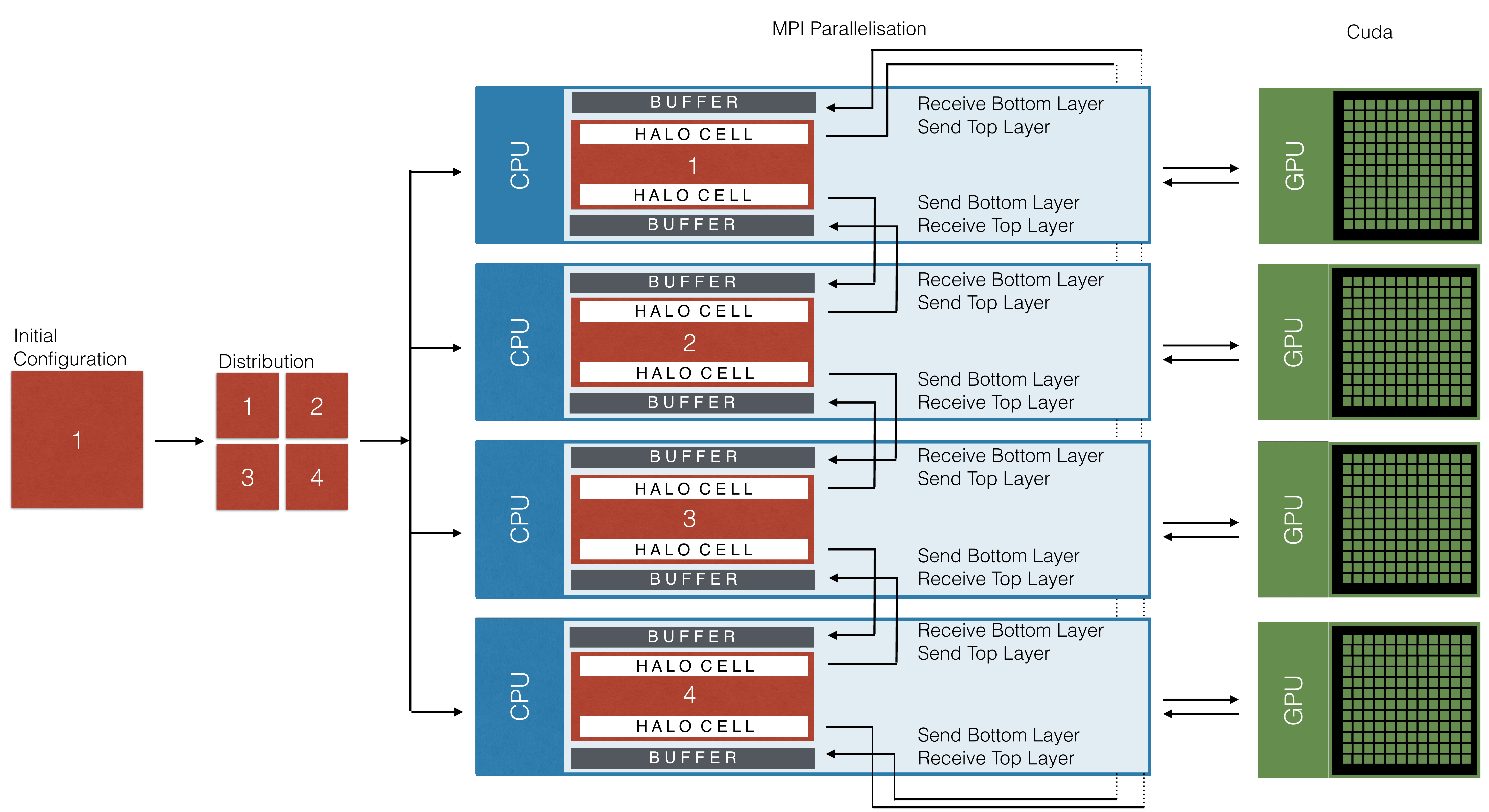}
	\caption{Flowchart outlining SMAUG+ with the implemented MPI parallelisation technique. The red boxes demonstrate the initial and distributed model configurations. The configuration is equally divided and spread around the different CPUs and GPUs, using MPI and CUDA. The white rectangles show the 'halo' cells, the grey rectangles demonstrate the buffers for storing the exchanged information. The red domains within each process show the actual mesh outline and the blue boundaries mark data, which is stored by different processes. The numerically intensive calculations are sent and performed by the GPUs. The GPUs distribute the subdomains further for calculating the actual numbers. These calculations are performed by the thousands of GPU cores (green rectangles). The figure is an example of a $2 \times 2$ configuration.}
	\label{mpi}
\end{figure*}

Numerical simulations are one of the most important tools for studying astrophysical magnetohydrodynamic (or MHD) problems since the birth of computer science. MHD modelling the physical processes of a complex astrophysical observation frequently requires enormous computational efforts with high compute performance because of the often strong inhomogeneous and/or stratified magnetised plasma medium, just to name two challenges. Advances in modern processing unit technology allows us to solve more and more complex physical problems by using faster and higher number of central processing units (CPUs) or accelerators, such as graphical processing units (GPUs). Many implementations of MHD exist using a range of approaches and methods \citep[see, e.g., ][]{Stone1992a, Stone1992b, Stone1992c, Toth1996, arber2001staggered, keppens2003adaptive, hasan2005dynamics, shelyag2008magnetohydrodynamic, Stone2008}, however, numerical performance of GPUs is recently exploited for scientific purposes and it is becoming used by an increasing number of diverse studies for modelling MHD problems \citep[e.g.][]{Wong2009, Pang2010, griffiths2015abs}. 

Multi-GPU (mGPU) systems are able to provide further benefits, such as larger computational domains and substantial compute time savings, the latter resulting in saving of operational costs. Many studies demonstrate the performance effectivity for solving various astrophysical problems with mGPU architecture. For instance, \cite{Schive2011} used a parallel GPU-accelerated adaptive-mash-refinement method for solving hydrodynamic problems. The mGPU systems allow us to achieve orders of magnitude performance speed-up compered to CPU cores \citep{wong2014global, wong2014efficient}. However, the mGPU systems also enable to extend considerably the investigated model size or increase the resolution of the computational domain, therefore allowing to obtain more  details.

The Sheffield Advanced Code (SAC) is a fully non-linear MHD numerical tool for simulating, in particular, linear and non-linear wave propagation in strongly magnetised plasma with structuring and stratification \citep{shelyag2008magnetohydrodynamic}. SAC is  based on the seminal Versatile Advection Code (VAC) \citep[see][]{Toth1996}. \cite{griffiths2015abs} developed the parallel version of SAC using a single GPU. \cite{griffiths2015abs} reported significant speed up, by demonstrating simulations of the Orszag-Tang vortex on a single NVIDIA M2070 GPU, being 18 times faster than the same problem running on a single core of an Intel Xeon X5650 CPU.

Our primary aim, here,  is to introduce a fast MHD code for gravitationally highly stratified media by further developing SMAUG. Therefore we introduce SMAUG+, running it on a multi-GPU system for allowing increased computational domain, i.e. enabling larger  physical model size computations. The developed software provides the opportunity to execute/perform simulations of MHD wave propagation mimicking the strongly magnetised solar atmosphere, in particular, representing the lower solar atmosphere from photosphere to low corona. Such approach is important, as there are a number of high-resolution ground- (e.g. SST - Swedish Solar Telescope, La Palma; DKIST - Daniel K. Inouye Solar Telescope, USA to be commissioned in 2019 or the EST - European Solar Telescope, to be realised by the second half of the next decade) and space-based (e.g. Hinode, SDO - Solar Dynamics Observatory, IRIS - Interface Region Imaging Spectrograph) facilities providing a wealth of earlier unforeseen observational details that need now to be understood.

\section{Numerical Approach}

%\subsection{Equations and methods}

The SMAUG+ is a numerical finite element solver, which is based on addressing the ideal fully non-linear 3-dimensional MHD equations. In particular, it is suitable to model linear and non-linear wave propagation in strongly magnetised plasma with structuring and stratification. The code is designed to solve multi-dimensional hyperbolic systems of partial differential equations. The governing equations of compressible MHD with stratification in their conservative form are:

\allowdisplaybreaks
\begin{align*}
	\frac{\partial \rho}{\partial t} + \nabla \cdot ( \bm{v} \rho )   &=  0, \\ 
	\frac{\partial (\rho \bm{v})}{\partial t} + \nabla \cdot (  \bm{v} \rho   \bm{v} -  \bm{B} \bm{B}) + \nabla p_{t}  &= \rho g, \\ 
	\frac{\partial e}{\partial t} + \nabla \cdot (  \bm{v}e - \bm{B} \bm{B} \cdot  \bm{v} +  \bm{v}p_{t}) + \nabla p_{t}  &= \rho  \bm{g} \cdot  \bm{v}, \\ 
	\frac{\partial  \bm{B}}{\partial t} + \nabla \cdot (  \bm{v} \bm{B} -  \bm{B} \bm{v})  &=  0, 
\end{align*}

where $\rho$ is the density, $v$ is the velocity, $B$ is the magnetic field, $e$ is the energy density and $g$ is the gravity. The total pressure  $p_{t}$ is defined by: 

\begin{equation*}
	p_{t} = p_{k} + \frac{ \bm{B}^{2}}{2}, 
\end{equation*}

\noindent
where, $p_{k}$ is the kinetic pressure:

\begin{equation*}
	p_{k} = (\gamma - 1) \bigg( e - \frac{\rho \bm{v}^2}{2} - \frac{\bm{B}^2}{2} \bigg).
\end{equation*}

Fourth-order central differencing method is applied for solving the spatial derivatives and fourth-order Runge-Kutta solver is performed for solving the time derivatives. By virtue of their symmetry, central differencing schemes are conservative, with the desired side effect that the solver conserves the divergence of the magnetic field. We employ hyper-resistivity and hyper-diffusion for increasing the numerical stability of the calculated MHD equation, based on the implementation by e.g., \cite{caunt20013d, stein1998simulations, shelyag2009acoustic}. The hyper-viscosity coefficient calculations require the estimation of maximum wave speed in the computational domain.

By applying the central difference approximation to the hyperbolic differential governing equations, the solutions are unstable with a spurious oscillatory behaviour. Hence, numerical diffusion is applying for stabilising the code. The primary purpose of the diffusion terms is to compensate for the anti-diffusion from truncation errors arising in the computation of temporal and spatial derivatives. When the diffusion is correctly tuned, the resulting evolution is non-diffusive. In addition, the diffusion terms control the steepness of shocks by becoming large wherever the compression is large.

The MHD equations and the hyper-diffusion source terms are described in details in \cite{griffiths2015abs}. Numerical instabilities could be generated by the applied central differencing method, hence obtaining the solutions of the shocked systems could be difficult. The hyper-viscosity parameter is the ratio of the forward difference of a parameter to third order and first order. The temporal evolution of the hyper-viscosity term helps to identify numerical noise. If it is necessary, the hyper-viscosity term also helps to smooth the identified noise. For further details of the algorithm and the tests the reader is referred to the papers describing The Sheffield Advanced Code (SAC) \citep{Fedun2011c, Fedun2011a, Fedun2011b} and SMAUG, the parallel version of the SAC employing a single GPU \citep{griffiths2015abs}.

\section{Parallelisation Techniques}

MPI and NVIDIA CUDA allow us to develop a hybrid environment with multiple GPUs and multi-core CPUs. The MPI provides a message-passing parallel programming model: data is moved from the address space of one process to that of another process through cooperative operations on each process. We use the MPI methods to spread the initial model configuration. 

Figure \ref{mpi} shows the principals of an example system of architecture. The red boxes represent the computational domain of the initial model configuration (e.g. for the Orszag-Tang vortex problem). The original grid, however, is now divided into four equal sub-regions, as indicated by the successive serial numbers. Each sub-region is assigned to a CPU. The CPUs are able to communicate with each other using communication fabrics, such as MPI messaging, OMNI-Path technology \citep{birrittella2015intel}. 

Figure \ref{mpi} sketches an example of the CPU-CPU MPI communication. Exchanging information between the sub-domains with 'halo' layers is a common practice in parallel computation on CPUs \citep{kjolstad2010ghost}. The halo layers are demonstrated by the white rectangle within the computation domains (red rectangles). The data are obtained from (or sent to) the buffer of the top (or bottom) neighbour processors (indicated by grey rectangles). By sending and receiving only 'halo' cells and not the full grid, we reduce the size of the communications. Based on halo messaging technique employed in the SMAUG and SAC code, we demonstrate an example code of the implementation of MPI messaging.

\begin{lstlisting}
void exchange_halo(vector v)
	{
		//gather halo data from v into gpu_buffer1
		cudaMemcpy(host_buffer1, gpu_buffer1, ...);
		MPI_Isend(host_buffer1, ..., destination, ...);
		MPI_Irecv(host_buffer2, ..., source, ...);
		MPI_Waitall(...);
		cudaMemcpy(gpu_buffer2, host_buffer2, ...);
		//scatter halo data from gpu_buffer2 to 
		//halo regions in v
	}
\end{lstlisting}

The CUDA platform, however, provides us to access GPU accelerated solutions. Here, the actual numerical methods are performed/applied by the GPUs. With MPI and CUDA we can use multiple GPUs simultaneously for e.g., increasing model resolution or archiving speed-up. 

\section{The HPC facility}

We ran the simulations using two different HPC facilities, namely, the University of Cambridge (Wilkes) and the University of Sheffield (ShARC) architectures.  The Wilkes computer is based on Dell T620 mashines. Each Dell T620 has two NVIDIA Tesla K20c (5 GiB) card and 12 cores Intel Ivy Bridge 2.6 GHz CPU and 64 GByte memory. Overall, the cluster consists of 128 Dell nodes with 256 GPUs. Each Tesla K20c contains 2496 GPU cores, hence the total number of GPU cores is 638976 and the total number of CPU cores is 1536.

The ShARC cluster is based on Dell PowerEdge C4130 units with 16 cores Intel Xeon E5-2630 v3 (2.4GHz) processors and 64 GByte memory. Accessing the GPU nodes provides 8 NVIDIA Tesla K80 (24 GiB) graphical units. The total number of GPU cores is 39936.

\section{Verification and Validation}
\subsection{The Orszag-Tang vortex}

\begin{figure}
\centering
  \begin{tabular}{@{}cccc@{}}
    \includegraphics[width=95mm]{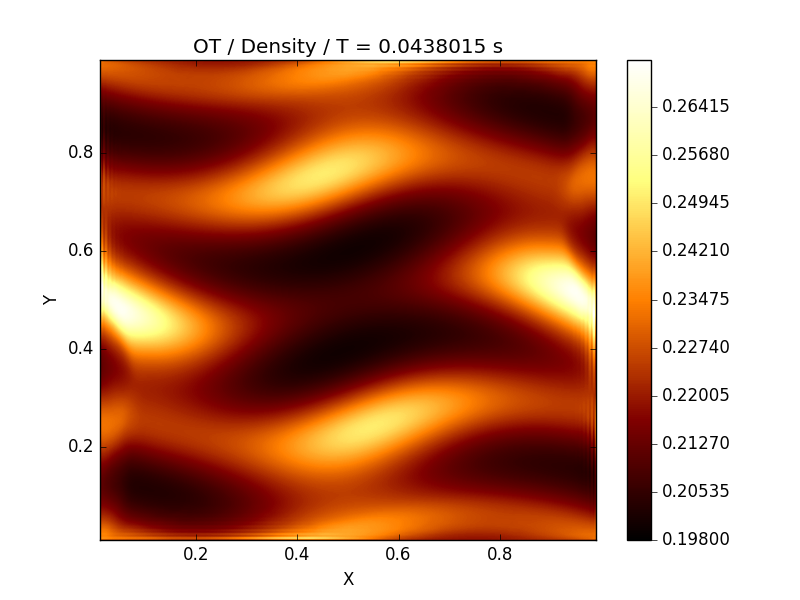} \\
    \includegraphics[width=95mm]{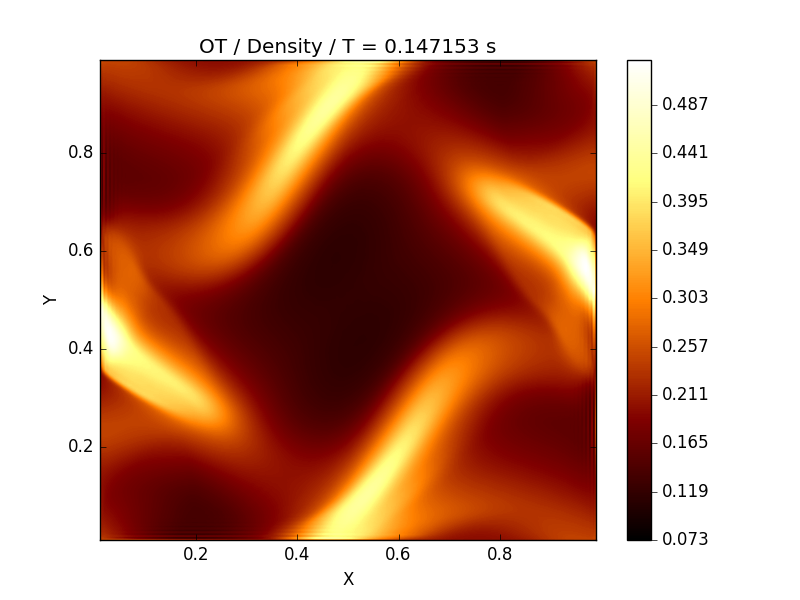} \\ 
    \includegraphics[width=95mm]{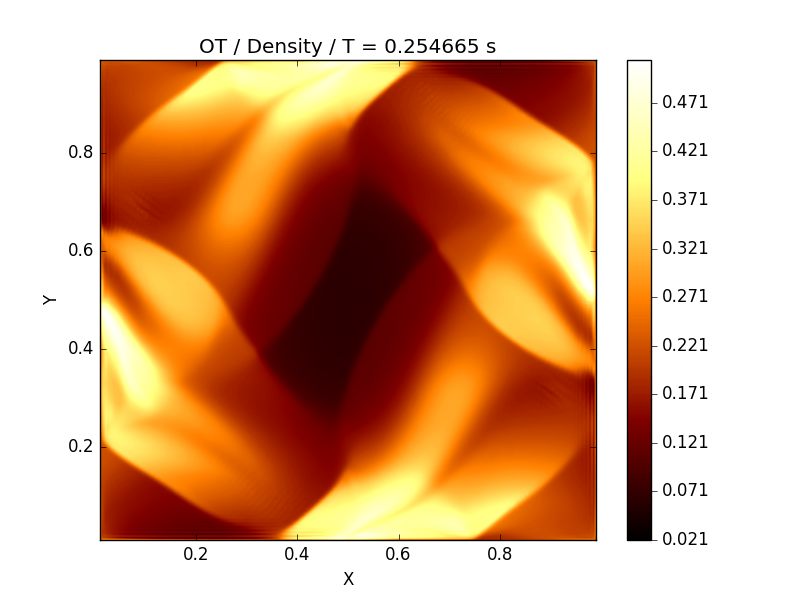}
  \end{tabular}
  \caption{Orszag-Tang vortex results computed by SMAUG+. The initial configuration contains $1000  \times 1000$ data and is distributed among 4 GPUs ($2  \times 2$). The figure shows the temporal variation of the density for $t_{1} = 0.04$ s, $t_{2} = 0.14$ s and $t_{3} = 0.25$ s. The simulation ran on the ShARC cluster.}
  \label{ot}
\end{figure}

The single GPU code (SMAUG) is already validated using e.g. the Brio-Wu shock tube test \citep{griffiths2015abs} for 1-dimensional MHD. The Orszag-Tang vortex is a common validation test employing two-dimensional non-linear MHD. This test is able to verify that the code employing it can handle MHD shock formations and shock interactions efficiently and cost-effectively. The horizontal and vertical non-dimensional domain sizes are set to $0 \leq x \leq 1; 0 \leq y \leq1$, the adiabatic index $\gamma = 5/3$, the gas pressure and the density are constants. The boundary conditions are set to periodic at each boundary. The magnetic field components ($B_{x}$, $B_{y}$) and the initial velocity components ($V_{x}$, $V_{y}$) are defined by:

\begin{align*}
	B_{x}  &=  - B_{0} \sin(4 \pi y), \\ 
	B_{y}  &=  B_{0} \sin(2 \pi x), \\ 
	V_{x}  &=   \sin(2 \pi y), \\ 
	V_{y}  &=   \sin(2 \pi x). 
\end{align*}

\begin{table}[]
	\centering
	\caption{Timings for 100 iterations for the Orszag-Tang test. The timing results are based on the simulations performed at the Cambridge Wilkes Cluster.}
	\label{sim}
	\begin{tabular}{llll}
		\hline
		Grid Size        &  Number                  & Time {[}s{]} & Time {[}s{]} \\
							   &	GPUs			          & Without HD & With HD     \\
		\hline
		$1000 \times 1000$ & $1 \times 1$   & 34.50   & - \\
		$1000 \times 1000$ & $2 \times 2$   & 11.19   & - \\
		$1000 \times 1000$ & $4 \times 4$   & 13.70   & - \\
		$2044 \times 2044$ & $2 \times 2$   & 41.32   & 184.10 \\
		$2044 \times 2044$ & $4 \times 4$   & 43.39   & 199.89 \\
		$4000 \times 4000$ & $4 \times 4$   & 77.44   & 360.71 \\
		$8000 \times 8000$ & $8 \times 8$   & 61.70   & 253.80 \\
		$8000 \times 8000$ & $10 \times 10$ & 41.00 & 163.60 \\
	\end{tabular}
\end{table}

We ran a series of simulations with different simulation box resolution as seen in the Table \ref{sim}. The timings with and without hyper-diffusion (HD) are distinguished, because we would like to understand the impact of the different code segments on the communication overhead between the GPUs.

Figure \ref{ot} is a set of snapshots of an example Orszag-Tang simulation. The panels shows the temporal variation of the density on a linear colormap. As the bottom panel of Figure \ref{ot} shows, various string waves pass through each other. This motion creates turbulent flow in different spatial scales. Figure \ref{ot} demonstrates that there is a convincing agreement between the results of SMAUG+ and their counterpart output of SMAUG \citep{griffiths2015abs} and SAC \citep{shelyag2008magnetohydrodynamic}.

\subsection{The Brio-Wu shock tube test}

The Brio and Wu shock tube is an excellent test for solving a Riemann-type problem \citep{roe1981approximate}. The shock tube is a 1D ideal MHD test problem in which the initial conditions of the model feature a discontinuity in the centre of the configuration, i.e., the left and right states are initialised with different values \citep{brio1988upwind}. On either side of the discontinuity, the initial parameters are: $p_{l},\rho_{l},B_{yl}  = 1, p_{r}  =  0.1, \rho_{r}  =  0.125, B_{yr}  =  -1$ and $B_{x}  =  0.75$. For the method employed by \cite{brio1988upwind}, the exact solution is approximated by a linearised version, averaged on either side of the discontinuity.

\begin{figure}
	\centering
	\hspace*{-0.2cm}\includegraphics[width=85mm]{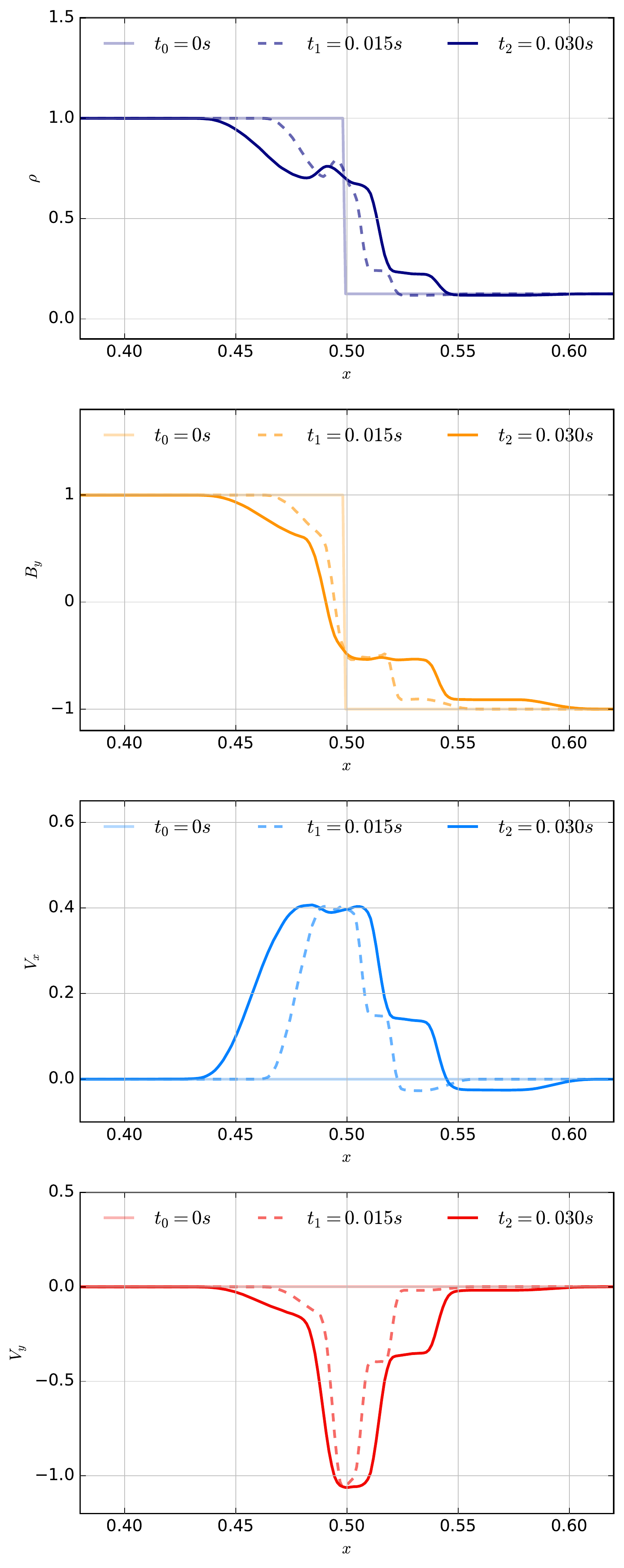}
	\caption{Numerical solution of Brio and Wu shock tube problem numerical solution taken at time $t_{0}=0$, $t_{1}=0.015$ and $t_{2}=0.03$. The density (purple colour), the tangential magnetic field (yellow colour), the tangential velocity (blue colour) and the (red colour) normal velocity are distinguished among the panels. The initial configuration was equally distributed among 4 GPUs.}
	\label{bw}
\end{figure}

Figure \ref{bw} shows various features confirming the accurate run on the code. The slow compound wave, the contact discontinuity, the slow shock and the fast rarefaction can be seen. Running the problem on a numerical domain with 800 grid points, gives an excellent agreement with the original SAC  \citep{shelyag2008magnetohydrodynamic} and SMAUG \citep{griffiths2015abs} results. We distributed the initial configuration among up to 10 GPUs. The tests demonstrate that code is sufficiently robust so that it can handle supersonic MHD turbulence.

\section{Parallel Performance}

The actual parallel performance of the applied models is determinated by various factors, such as: (i) the granularity of the parallelizable tasks, (ii) the communication overhead between the nodes, (iii) finally, the load balancing. The load balancing refers to the distribution of the data among the nodes. If the data distribution is not balanced, some of the GPUs with less load must wait until the heavily loaded GPUs finish the job. We always use equally divided configurations, hence all the GPUs are equally loaded. The granularity of the parallel task represents the amount of work that will be carried out by a certain node. The communication overhead is the cost of sending and receiving information between the different nodes. In our case, the overhead is built-up by two components: the MPI node communication and the CPU-GPU information transfer, namely the functions $cudaMemcpy()$, $MPI{\char`_}Isend()$, $MPI{\char`_}Irecv()$ and $MPI{\char`_}Waitall()$. The actual running time $T$ of a parallel job task is,

\begin{equation*}
 	T = T_{comp} + T_{comm} + T_{idle},
\end{equation*}
 
where, $T_{comp}$ is the useful computation time, $T_{comm}$ is the communication overhead and $T_{idle}$ is the waiting time until all the GPUs finishes a certain step. The $T_{idle}$ could be minimised by optimal load balance and $T_{comm}$ could be relatively small by choosing optimal granularity.

Let us define the quantity $\tau$ for characterising the parallelisation overheads:

 \begin{equation*}
 	\tau = \ln \frac{T_{comp}}{T_{comm} + T_{idle}}.
 \end{equation*}
 
\begin{figure}
	\centering
	\hspace*{-0.2cm}\includegraphics[width=91mm]{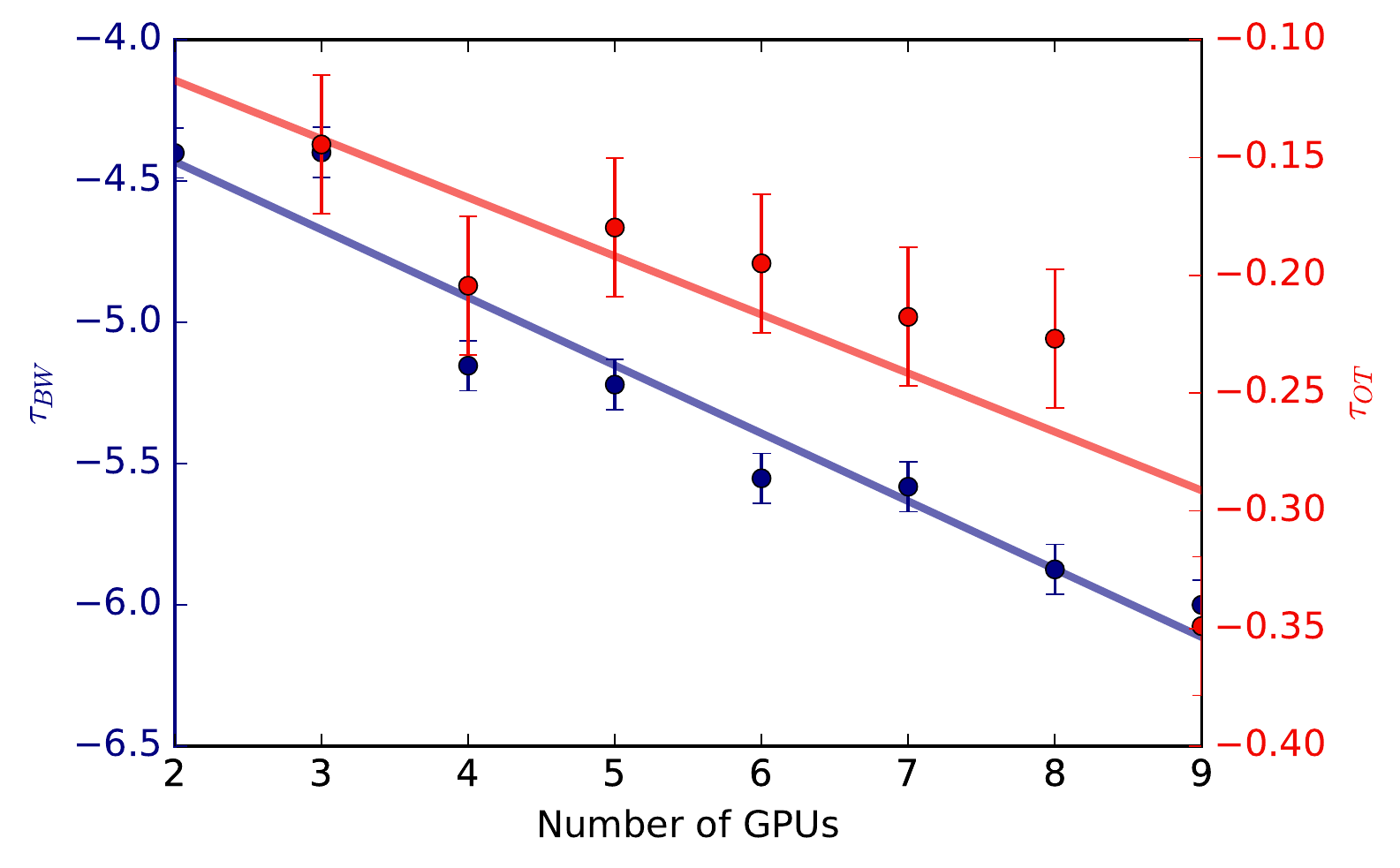}
	\caption{Brio-Wu and Orszag-Tang communication overhead test. The vertical axis represents the parameter $\tau$, the ratio of useful calculation and communication. The horizontal axis shows the number of applied GPUs. The size of initial configuration is constants (800 data points for Brio-Wu and $2044 \times 2044$ for the Orszag-Tang test. The increasing parallelism alone cannot increase performance.}
	\label{slowdown}
\end{figure}

Parallel slowdown ($T_{comm} + T_{idle} > T_{comp}$) could be the result of a communication bottleneck. More GPUs must spend more time for communication. Sometimes, the communication between the GPUs takes actually more time than the useful calculation. This situation is demonstrated by the Figure \ref{slowdown}. The blue values represent Brio-Wu simulations, using up to 10 GPUs. As the number of GPUs increases, the parameter $\tau$ decreases. We kept the model size constant. The range of the parameter $\tau$ shows extremely small numbers, e.g $\tau= -4.5$ means only $1\%$ of the running time was actual calculation. This value is extremely small, because it is likely that much of the communications overhead arises from routines, used for transferring data within the GPU memory. From the GPU memory the data must be transferred to the system memory. From here, the CPU will send the information to another CPU node, finally this node transfers the data to the GPU and so on. This continuous data transfer significantly jeopardise the parallel performance. This is the consequence of using not computationally dense GPUs. The red values in Figure \ref{slowdown} shows similar behaviour but the used model is a $2044 \times 2044$ Orszag-Tang configuration. In this case, a GPU calculates a larger amount of data, hence the nodes are more computationally dense. Around $30\%$ of the running time is the actual calculation. The total running time of the 9-GPU configurations in both cases is around 4 times slower than the 2-GPUs configuration.
 
As shown above, choosing the non-optimal configuration could cause massive wasting of computational power. To avoid parallel slowdown the following must be considered: (i) Only increasing parallelism will not provide the best performance \citep{chen1990impact}. Increased parallelism with non-optimal data granularity could easily cause parallel slowdown, as seen above. (ii) The amount of exchanged MPI messages must be reduced as much as possible for the best performance \citep{thakur2004minimizing}. It also means that a single GPU could give better performance than multiple GPUs, if the applied model size is the same. (iii) Task must be enough to overleap the parallel communication overheads. The processes must have a higher task granularity if the number of applied GPUs increases. For avoiding the communications overhead, it is advisable to use always arithmetically dense GPUs. The adjustment of the model size and the decomposition of the initial configuration must be optimal. (iv) It is possible to improve communication performance by using higher-performance communication hardware, but it may turn out expensive. The collective communication, can also improve communication performance, since it optimises the communication based on the hardware, network and topology.

By applying the above principle parallel performance speed-up is possible. In Table \ref{sim}, the $1000 \times 1000$ Orszag-Tang test with 4 GPUs is around 3 times faster than the 1 GPU configuration. The $8000 \times 8000$ test shows 1.5 times speed-up between 64 and 100 GPUs.  The simulations with HD show similar properties than the simulations without HD. The timings with HD are significantly longer than without HD which shows that the HD functions are extremely computation intense segments of the SMAUG+.

\section{Discussion}

Future, high-performance computer architectures are enabling an increasing use of large numbers of GPU accelerators. It is crucial to develop codes enabling an the available fast communication between the GPUs. Our approach to exchange information between decomposed domains not only occupies one core in each CPU but also it may become a bottleneck to speed and scaling because of indirect data transfer between GPUs. This issue may be solved by direct data transfer between GPUs introduced by the GPU Direct technology.

However, the primary aim of our approach is to archive extended model size. A single GPU is only able to support limited memory but using our method an mGPU system provides as much memory as the total of GPUs. The only disadvantage is the communication overhead, however, an mGPU system may still be faster and significantly cheaper than a multi-CPU approach. For the same price, a GPU contains orders of magnitude more processing cores than a CPU. Our approach provides affordable desktop high-performance computing.

The challenge with advancing these codes is the identification and application of the current communication standards: different possibilities including NVLINK and various versions of the GPU Direct technology. These implementations are sensitive to the system configuration, making it challenging to successfully implement the code.

\section{Conclusion}

We have demonstrated that SMAUG+ is able to successfully compute numerically valid and large MHD problems by distributing the needed compute tasks across multiple GPUs. We conclude that the speed-up of the MHD simulations performed depends on the distribution of the initial architectural configuration. In some cases, the running time could be slower if unwanted communication overhead arises between the higher number of GPU. However, and we propose this is a key point, a not over-distributed simulation could show significant speed-up. Further performance enhancements are also feasible through application architecture modification. By using our method an mGPU system essentially is able to  provides as much memory as the total of applied graphical accelerators offer. The algorithm has been implemented in 3D testing of 3D models that will be completed over a forthcoming projects.

\section*{Acknowledgments}

MG and RE are grateful to STFC (UK), and RE acknowledges The Royal Society (UK) for the support received. This research was made use of SunPy, an open-source and free community-developed Python solar data analysis package \citep{mumford2013sunpy}. The authors are grateful to the University of Cambridge HPC service for providing access to the Wilkes cluster and to Filippo SPIGA,  Head of Research Software Engineering (RSE) University of Cambridge for providing assistance with running the the SMAUG+ code on the Wilkes cluster.

\section*{References}

\bibliographystyle{apalike}

\bibliography{bibfile}

\end{document}